\documentclass[12pt,preprint]{aastex}
\newcommand{\mjupe}{M$_{\rm JUP}$}
\newcommand{\msini}{\ensuremath{M \sin{i}}}
\newcommand{\feh}{\ensuremath{[\mbox{Fe}/\mbox{H}]}}
\newcommand{\rphk}{\ensuremath{R'_{\mbox{\scriptsize HK}}}}
\newcommand{\mv}{\ensuremath{M_{\mbox{\scriptsize V}}}}

\newcommand{\dmv}{\ensuremath{\Delta\mv}}
\newcommand{\teff}{\ensuremath{T_{\mbox{\scriptsize eff}}}}
\newcommand{\persec}{\ensuremath{\mbox{s}^{-1}}}
\newcommand{\mps}{\mbox{m} \persec}
\newcommand{\mjup}{\ensuremath{\mbox{M}_{\mbox{Jup}}}}

\newcommand{\Mjupsmall}{\mbox{\scriptsize M}_{\mbox{\tiny Jup}}}
\def\astrosun {\mbox{$\odot$}}
\newcommand{\Msol}{\ensuremath{\mbox{M}_{\astrosun}}}

\shorttitle{Catalog of Nearby Exoplanets}
\shortauthors{Butler et al.}

\begin{document}
\title{Catalog of Nearby Exoplanets\altaffilmark{1}}
\altaffiltext{1}{Based on observations obtained
at the W. M. Keck Observatory, which is operated jointly by the
University of California and the California Institute of Technology.
The Keck Observatory was made possible by the generous financial
support of the W. M. Keck Foundation.}
\author{R. P. Butler\altaffilmark{2}, J. T. Wright\altaffilmark{3},
 G. W. Marcy\altaffilmark{3,4}, D. A Fischer\altaffilmark{3,4}, 
 S. S. Vogt\altaffilmark{5}, C. G. Tinney\altaffilmark{6}, 
 H. R. A. Jones\altaffilmark{7}, B. D. Carter\altaffilmark{8}, 
 J. A. Johnson\altaffilmark{3}, C. McCarthy\altaffilmark{2,4}, 
 A. J. Penny\altaffilmark{9,10}}

\altaffiltext{2}{Department of Terrestrial Magnetism, Carnegie Institute of Washington, 5241 Broad Branch Road NW, Washington, DC 20015-1305}
\altaffiltext{3}{Department of Astronomy, 601 Campbell Hall, University of California, Berkeley, CA 94720-3411}
\altaffiltext{4}{Department of Physics and Astronomy, San Francisco State University, San Francisco, CA 94132}
\altaffiltext{5}{UCO/Lick Observatory, University of California, Santa Cruz, CA 95064}
\altaffiltext{6}{Anglo-Australian Observatory, PO Box 296, Epping. 1710. Australia}
\altaffiltext{7}{Centre for Astrophysics Research, University of
  Hertfordshire, Hatfield, AL 10 9AB, England}
\altaffiltext{8}{Faculty of Sciences, University of Southern Queensland, Toowoomba. 4350. Australia}
\altaffiltext{9}{Rutherford Appleton Laboratory, Chilton, Didcot, Oxon
  OX11 0QX, UK}
\altaffiltext{10}{SETI Institute, 515 N. Whisman Road, Mountain View,
  CA 94043}

\begin{abstract}
We present a catalog of nearby exoplanets.  It contains the 172 known
low-mass companions with orbits established through radial velocity
and transit measurements around stars within 200 pc.  We include 5
previously unpublished exoplanets orbiting the stars HD 11964, HD
66428, HD 99109, HD 107148, and HD 164922.  We update 
orbits for 90 additional exoplanets including many whose orbits have not 
been revised since their announcement, and include radial
velocity time series from the Lick, Keck, and  
Anglo-Australian Observatory planet searches.
Both these new and previously published velocities are more precise
here due to improvements in our data reduction pipeline, which we
applied to archival spectra.  We present a brief summary of the global
properties of the known exoplanets, including their distributions of orbital
semimajor axis, minimum mass, and orbital eccentricity.
\end{abstract}


\section{Introduction}

It has now been more than 10 years since the discovery of the first
objects that were identified as planets orbiting normal stars.  The epochal
announcement in 1995 October of 51 Peg b \citep{Mayor_queloz} was
confirmed within a week \citep{Marcy_51peg} and followed within 2
months by two other planets --- 47~UMa~b and 70~Vir~b
\citep{Butler96a,Marcy_70vir}.  The unexpected diversity and mass
distribution of exoplanets was represented well by those first three
planets, as the first one orbits close-in, the second orbits beyond 2 AU, and the last
resides in a very eccentric orbit.  The paucity of
companions having larger masses, with \msini\ between 10-80 \mjupe,
suggested a mass distribution separated from that of stars, rising
with decreasing mass and peaking below 1 \mjup \citep{Marcy_Butler00,
Halbwachs00, Udry03b}.

During the past 10 years, over 160 exoplanet candidiates have been identified
orbiting stars within 200 pc, and most have been detected by Doppler search
programs based at the Keck, Lick, and
Anglo-Australian Observatories \citep[the California \& Carnegie and
  Anglo-Australian planet searches, e.g.][]{Butler96b, Tinney01} and
teams based at l'Observatoire de Haute Provence and La Silla
Observatory \citep[the Geneva Extrasolar Planet Search, e.g.][]{Mayor03}.  Other Doppler programs have contributed
important discoveries of nearby planets \citep{Cochran97, Endl03,
Noyes99, Kurster03, Charbonneau00, Sato05}.  One nearby planet,
TrES-1, has been discovered by its transit across the
star \citep{Alonso04}. 

Here we present a catalog of all known exoplanets that reside within 200 pc,
containing the vast majority of well-studied exoplanets.  This
distance threshold serves several purposes.  First, nearby planets and their
host stars are amenable to confirmation and follow-up by a variety of
techniques, including high resolution imaging and stellar spectroscopy
with high signal-to-noise ratios, and astrometric follow up
\citep[e.g.][]{Benedict02, McArthur04}.  In addition, milliarcsecond
astrometry for planet-host stars within 200 pc can provide precise distance
estimates, and most planet search target stars within 100 pc already
have parallaxes from  Hipparcos \citep{PerrymanESA}.  Thirdly,
nearby planet-host stars are bright enough to permit precise
photometric and chromospheric monitoring by telescopes of modest size,
permitting careful assessment of velocity jitter, starspots, and
possible transits, e.g., \citet{HenryG99, hetal00a, Queloz166435,
ehf03}.

This paper updates the last published list of exoplanets \citep{Butler02}.
The growth of the field is reflected by the discovery of
over 100 planets in the 3 years since the publication of that list of 57
exoplanets.

About a dozen exoplanet candidates have been discovered that reside
beyond 200 pc, 
including a half dozen in the Galactic bulge found in the OGLE survey
and a few other planets found by microlensing \citep[e.g.][]{Torres03,
  Konacki03, Bouchy05}.  Perhaps most notable are the first 
planets ever found outside our Solar System, orbiting a pulsar
\citep{Wolszczan92}.  Such distant planets reside beyond the scope of
this catalog. 

We include known companions with minimum masses (\msini) up to 24
\mjup.  This is well above the usual 13 \mjup\ deuterium-burning limit
for planets adopted by the IAU.  We do this for for two
reasons.  First, uncertainties in stellar mass and orbital 
inclination complicate the measurement of sufficiently precise masses
to apply a robust 13 \mjup\ cutoff.
Secondly, there is little or no evidence indicating that such a
cutoff has any relevance to the formation mechanisms of these
objects.  We therefore use a generous minimum mass criterion for
inclusion in this catalog, and decline to choose a precise definition
of an ``exoplanet''.  

Two other planet candidates were detected by direct imaging, 2M1207 b
\citep{Chauvin04}, and GQ Lup b \citep{Neuhauser05}.  We exclude these
from the tabular catalog due to their considerably uncertain orbital
periods, eccentricities, and masses.  Similarly, we exclude many
Doppler-detected planets due to their lack of data spanning one full
period, which precludes a secure determination of their orbits and
minimum masses.

One might question the value of a catalog of exoplanets in the face of
such rapid discovery.  Without question, the catalog presented here
will become out of date before it is printed.\footnote{Note added in
  proof: Indeed, after submission this paper Lovis et al. (2006)
  announced a triple-Neptune system orbiting HD 69830, Hatzes et
  al. (2006) confirmed a 2.3 \mjup\ planet orbiting Pollux,
  J. T. Wright et al. (in preparation) announced a planet orbiting HD
  154345 and a second planet orbiting HIP 14810, and J. A. Johnson et
  al. (in preparation) announced a hot Jupiter orbiting HD 185269.
  Our group will maintain an up-to-date version of the Catalog of
  Nearby Exoplanets on the World Wide Web at http$://$exoplanets.org}
However, this catalog 
offers many attributes of unique value.  First, it contains updated
orbital parameters for 90 exoplanets, computed anew from our
large database of Doppler measurements of over 1300 stars from the
Lick, Anglo-Australian, and Keck Observatories obtained during the
past 18, 7, and 
8 years respectively \citep{Butler03, Marcy_Japan_05}.  These new orbital
parameters significantly supersede the previously quoted orbital
parameters in most cases.  

Second, we use the latest estimates of stellar mass to improve the
precision of the minimum planet mass, \msini
\citep{Valenti05}.  Thirdly, the catalog contains Doppler measurements 
for the planet host stars in our database, allowing 
both further analyses of these
velocities and novel combinations with other measurements.  The
publication of this archive foreshows a forthcoming work (Wright et
al., 2006, in prep) which will identify and catalog prospective
exoplanets and substellar companions of indeterminate mass and orbital period.
Finally,
the catalog will serve as an archive of known nearby exoplanets and
their parameters circa 2005.  The catalog may serve ongoing exoplanet
research, both observation and theory, and provide useful information
for future exoplanet studies of nearby stars.

\section{Data}
The radial velocity data here come from three sources:  observations
at Lick Observatory using the Hamilton spectrograph \citep{Vogt87}, at
Keck Observatory using HIRES \citep{Vogt94}, and at the
3.9 m Anglo-Australian Telescope using UCLES \citep{Diego90}.  These
instruments, their characteristics, and typical uncertainties in the
radial velocities they produce are discussed in the discovery papers
of the exoplanets planets found with them \citep[in
  particular][]{Fischer99, Butler98,Tinney01}.  We
explicitly note here upgrades over the years which have significantly
improved their precision at typical exposure times:  the Hamilton
spectrograph was upgraded in November 1994, increasing the precision of a
typical observation from $10-15$ m\persec\ to $\sim 4$ m\persec.   In
August 2004 HIRES was upgraded, increasing the precision of a typical
observation from $\sim 3$ m\persec to $\sim 1$ m\persec.  The
precision of UCLES data is $2 - 3$ m\persec.  

We have also revised our entire reduction pipeline, including an
overhauled raw reduction package which includes corrections for cosmic
rays and an improved flat-fielding algorithm, a more accurate
barycentric velocity correction which includes proper-motion
corrections, and a refined precision velocity reduction package
which includes a telluric filter and a more sophisticated deconvolution
algorithm.  We also now correct 
for the very slight non-linearity in the new HIRES CCD.  We have 
improved the characterization of the charge transfer inefficiency in
the old CCD which limited its precision to $\sim 3$ m\persec, a
problem not present in the new chip.

In previous works we have subtracted a constant
velocity such that the median velocity of the set was zero (since
these are differential measurements, one may always add an arbitrary
constant to the entire set).  Here, we have applied an offset to the
data so that the 
published orbital solution has $\gamma = 0$, where $\gamma$ is the
radial velocity of the center of mass of the system.  

For the above reasons the measurements listed here are more precise and
accurate than the values given in our previous publications, and will
not exactly match the values given in those works. There may
also be slight differences in the binning of measurements made within
about two hours of one other.

\section{Radial Velocities}

In the table of radial velocities for our stars (available in the
electronic edition of the {\it Journal}), we report the time of
observation, measured radial velocity, and formal uncertainty in each
measurement.  The uncertainties reported are measured from the
distribution of velocities measured from each of 400 parts of each
spectroscopic observation, as discussed in previous works
\citep[e.g.][]{Marcy05}, and do not include jitter.  We present a
sample of this data set in Table~\ref{vels}.
\clearpage
\begin{deluxetable}{lrrrc}
\tablecolumns{5}
\tablewidth{0pc} 
\tablecaption{\label{vels} Radial Velocities for Planet Bearing Stars}
\tablehead{ 
\colhead{Star Name} & \colhead{Time}& \colhead{Velocity} &
\colhead{Uncertainty} & \colhead{Observatory} \\
\colhead{} & \colhead{(JD-2440000)} & \colhead{(m \persec)} & \colhead{(m \persec)}&}

\startdata 
HD 2039 & 11118.057282 & 14.5 &8.5 & A\\
HD 2039 & 11118.960972 & -9 &15 & A\\
HD 2039 & 11119.944525 & 3 &11 & A\\
HD 2039 & 11121.038461 & 0 &14 & A\\
HD 2039 & 11211.951424 & -24 &16 & A\\
HD 2039 & 11212.923368 & -11 &11 & A\\
HD 2039 & 11213.974942 & -8 &15 & A\\
HD 2039 & 11214.917072 & -14 &10 & A\\
HD 2039 & 11386.322743 & -29 &15 & A\\
HD 2039 & 11387.298102 & -16 &11 & A\\
\enddata

\tablecomments{[The complete version of this table is in the electronic edition of
the Journal.  The printed edition contains only a sample.]}
\end{deluxetable}
\clearpage

The table contains five columns. The first contains the name of the
star.  The second contains the time of observation as a Julian
date.  The third contains the measured precision radial velocity at
that time, and the fourth, the uncertainty in this measurement.  The
final column contains a key indicating which observatory made the
observation:  'K' for Keck Observatory, 'A' for the AAT, and 'L' for Lick
Observatory. 

In addition to the uncertainties published here, there are known
sources of error associated with astrophysical jitter, the instrument,
and the analysis.  These sources combine to give an additional source
of noise, collectively termed ``jitter''.  The magnitude of the jitter
is a function of the spectral type of the star observed and the
instrument used.  \citet{Wright05} gives a model (for stars observed
before August 2004 at Keck) that estimates, to within a factor of
roughly 2, the
jitter for a star based upon a star's activity, color, \teff, and height
above the main sequence.  More recent
measurements on HIRES will have less jitter due to the improved characteristics
of the new CCD.  Nonetheless, we adopt this model as an additional
source of noise for all observations at all telescopes.  We report
these adopted jitter values in Table~\ref{proptable}

\section{Errors\label{error}}

We calculated uncertainties in orbital parameters through the following
method, described in \citet{Marcy05}:  We subtracted the best-fit orbital solution from
the data and interpreted the residuals as a population of random
deviates with a distribution characteristic of the noise in the data.  We
randomly selected deviates from this set, with replacement, and added
this ``noise'' to the velocities calculated from the best-fit solution
at the actual times of observation.  We then found the best-fit
orbital solutions to this mock data set.  Repeating this procedure 100
times, we produced 100 sets of orbital parameters. We report the
standard deviation of each individual parameter over the 100 trials
as the 1$\sigma$ errors listed in Table~\ref{orbittable}.  For the
derived quantities $a$ and \msini, we calculated these quantities
from each mock data set and report the standard deviation in those
quantities propagated with an assumed error of 10\% in the stellar
mass (which dominates the error budget for many planets).

Uncertainties in $e$ and $\omega$ become non-Gaussian when $\sigma_e
\gtrsim e/2.$; in particular $\omega$ and $\sigma_\omega$ become ill-defined
when $e=0$.  In order to report uncertainties in an intuitive manner, we
calculate $\sigma_e$ in such cases as the geometric mean of
$\sigma_{e\cos{\omega}}$ and $\sigma_{e\sin{\omega}}$.
In other words, for cases when $\sigma_e \gtrsim e/2.$, we effectively model the
uncertainties as a 2-d Gaussian in $(e\cos{\omega})$-$(e\sin{\omega})$-space where
the values of $e$ and $\omega$ reported in Table~\ref{orbittable} are
the coordinates of the center of the Gaussian, and the error in $e$ is
its width.

For succinctness, we express uncertainties using parenthetical notation, where the least
significant digit of the uncertainty, in parentheses, and that of the quantity
are understood to have the same place value.  Thus, ``$0.100(20)$'' indicates
``$0.100 \pm 0.020$'', ``$1.0(2.0)$'' indicates ``$1.0 \pm 2.0$'', and
``$1(20)$'' indicates ``$1 \pm 20$''. 

Spectroscopic parameters from SPOCS \citep{Valenti05}
have typical errors of 44 K in $T_{\mbox{eff}}$, 0.06 dex in $\log{g}$,
0.03 dex in \feh, and 0.5 km\persec\ in $v\sin{i}$.  Errors in
the corresponding parameters from \citet{Santos04a} and
\citet{Santos05} are 50 K, 0.12 dex and 0.05 dex, respectively
($v\sin{i}$ is not quoted in these sources).  We quote errors in parameters
from other sources explicitly.

\section{Stellar Properties}

Table~\ref{proptable} represents a compilation of data on the
properties of the host stars for the nearby exoplanets.  Columns 1-2
list the HD and Hipparcos numbers of the stars, and column 3 acts as
a gloss for stars with Flamsteed, Bayer, or commonly used Gliese designations
(e.g. 51 Peg, $\upsilon$ And, GJ 86), many of which appear in
Table~\ref{orbittable}.

Hipparcos \citep{ESA97} provides accurate distances and positions
to all stars in this catalog save two, BD -10\arcdeg3166 and the
host star of TrES-1.  We quote coordinates, \bv, $V$ magnitude, and
distance to stars from the Hipparcos catalog in columns 4-8.

Columns 9-13 contain \teff, $\log$ g, abundance, $v \sin{i}$, and
mass for these
stars, collected from the references listed in column 14.  Most of
these reported values come from the SPOCS catalog, 
\citep{Valenti05}, whose measurements are based on detailed
spectroscopic analysis and evolutionary models, and the catalogs of
\citet{Nordstrom04}, \citet{Santos04a} and \citet{Santos05}, 

Column 15 lists Mount Wilson $S$-values for many of the stars, most of
which are drawn from \citet{Wright04}, \citet{Tinney02}, and \citet{Jenkins05},
but some of which are new to this work, measured in the manner
described in \citet{Wright04}. Column 16 lists the height of the star above the main sequence,
\dmv\ \citep[a function of \mv\ and
  \bv\ defined in][]{Wright04b}.   Column 16 lists the jitter predicted by the model of \citet{Wright05} for those
stars for which we have updated orbital parameters.  We have added these
jitter values in quadrature to the formal uncertainties when fitting
for the orbital parameters listed in Table~\ref{orbittable}.  This
procedure is also discussed in \citet{Marcy05}.

\section{Catalog of Nearby Exoplanets}

Table~\ref{orbittable} presents the Catalog of Nearby Exoplanets.  For planets with recently published velocities and orbits
(e.g. the HD 190360 system in \citet{Vogt05}) or those for which we
have insufficient data for an orbital fit or no data at all (e.g. HD 1237 b), we quote the most recently published solution.  For
all others, the orbital parameters in Table~\ref{orbittable} 
represent the best-fit orbital solutions to the velocities in Table~\ref{vels}.  

The name of each host star appears once for each system of planets in
the first column.  Where
available, we use Bayer designations or Flamsteed numbers to identify
a star (e.g. 51 Peg, not HD 217014) since these
names are more mnemonic than HD and Hipparcos catalog numbers, which
are cross-referenced in Table~\ref{proptable}.  For stars with no HD number, (e.g. GJ 86), we use the most common designation in the literature.  The second
column gives the component name ({\it b}, {\it c}, etc.) of each
planet.  Component names are ostensibly assigned in order of
discovery.

Columns 3-9 report the parameters of a best-fit solution to the
observed radial velocities: {\it P}, the sidereal orbital period of the
planet in days; {\it  K} the semi-amplitude of the reflex motion of
the star in m \persec; {\it e}, the eccentricity of the planet's
orbit; $\omega$, the longitude of periastron of the planet's
orbit in degrees; $T_{\mbox{p}}$, the time of
periastron passage as a Julian Date; $T_{\mbox{t}}$, the mid-time of
transit assuming $i=90\arcdeg$; and the magnitude of a linear
trend (in m \persec) subtracted from the velocities required to
achieve the fit.  We have excluded $T_{\mbox{p}}$ values for those
fits where the eccentricity has been fixed at 0, except in cases
collected from the literature where $\omega = 0$ arbitrarily.  We have
not calculated $T_{\mbox{p}}$ or $T_{\mbox{t}}$ values for orbital
parameters collected from the literature, but we report them where
present.  Parameters for dynamical fits in Table~\ref{orbittable} from the
literature may use slightly different definitions of these parameters,
using Jacobi coordinates and synodic periods \citep[e. g.][]{Rivera05}.

Columns 10 and 11 contain the minimum mass (\msini) and orbital
radius ($a$) of the planet, calculated from the orbital
parameters and the mass of the host star ($M_{\star}$ given in
Table~\ref{proptable}) using the following definitions:

\begin{equation}
  \msini = K \sqrt{1-e^2} \left(\frac{P(M_{\star}+\msini)^2}{2\pi G}\right)^{1/3}
\end{equation}
\begin{equation}
  \left(\frac{a}{\mbox{AU}}\right)^3 = \left(\frac{M_{\star}+\msini}{M_{\astrosun}}\right)
  \left(\frac{P}{\mbox{yr}}\right)^2
\end{equation}

where $G$ is the gravitational constant.

Columns 11 and 12 report the quality of the fit as the r.m.s. of the
residuals and reduced chi-square $\chi^2_{\nu}$ for the appropriate number of degrees
of freedom., and
column 13 reports the number of observations used in the fit.  
Column 14 contains the reference for the quantities in columns 3-8,
11, 12, and 13.  For many planets (e.g. 51 Peg b), other groups have
published an orbital solution independent of ours.  In these cases,
we cite the most recent such solution parenthetically in column
14. When this independent solution is of comparable quality to that in
Table~\ref{orbittable}, we reproduce it in Table~\ref{alttable}.

\section{New exoplanets}
We announce here five new exoplanets, HD 11964 b, HD
99109 b, HD 66428 b, HD 107148 b, HD 164922 b.  Their orbital
parameters and 
the properties of their host stars are listed among the other entries
in the tables below.  The data for these detections were obtained at
Keck Observatory.  All of these exoplanets orbit inactive stars
($\log{\rphk} < -5$) which are metal-rich ($\feh > 0.1$). 

HD 11964 is somewhat evolved, sitting two magnitudes above the main
sequence.  The fit for HD 11964 b is good, but the 5.3 m \persec\
residuals are comparable to the 9 m \persec\ amplitude, making the
exoplanetary interpretation of the velocity variations somewhat in doubt.
\clearpage
\begin{figure}
\plotone{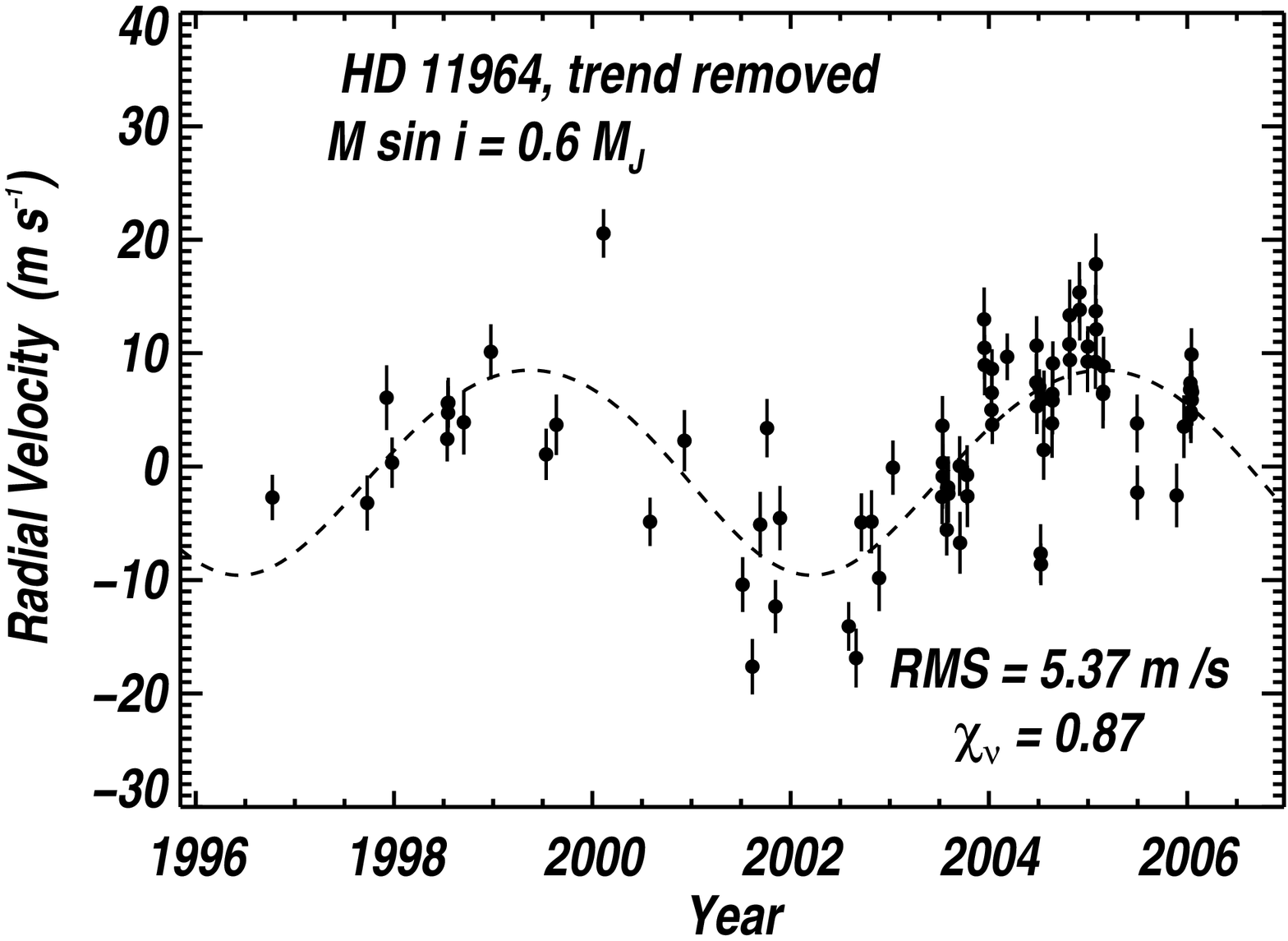}
\caption{\label{11964} Best-fit orbit to the radial velocities measured
  at Keck Observatory for HD 11964, with $P=5.8 yr$, $e \sim 0$,
  and $\msini =0.6 \mjup$.}
\end{figure}
\begin{figure}
\plotone{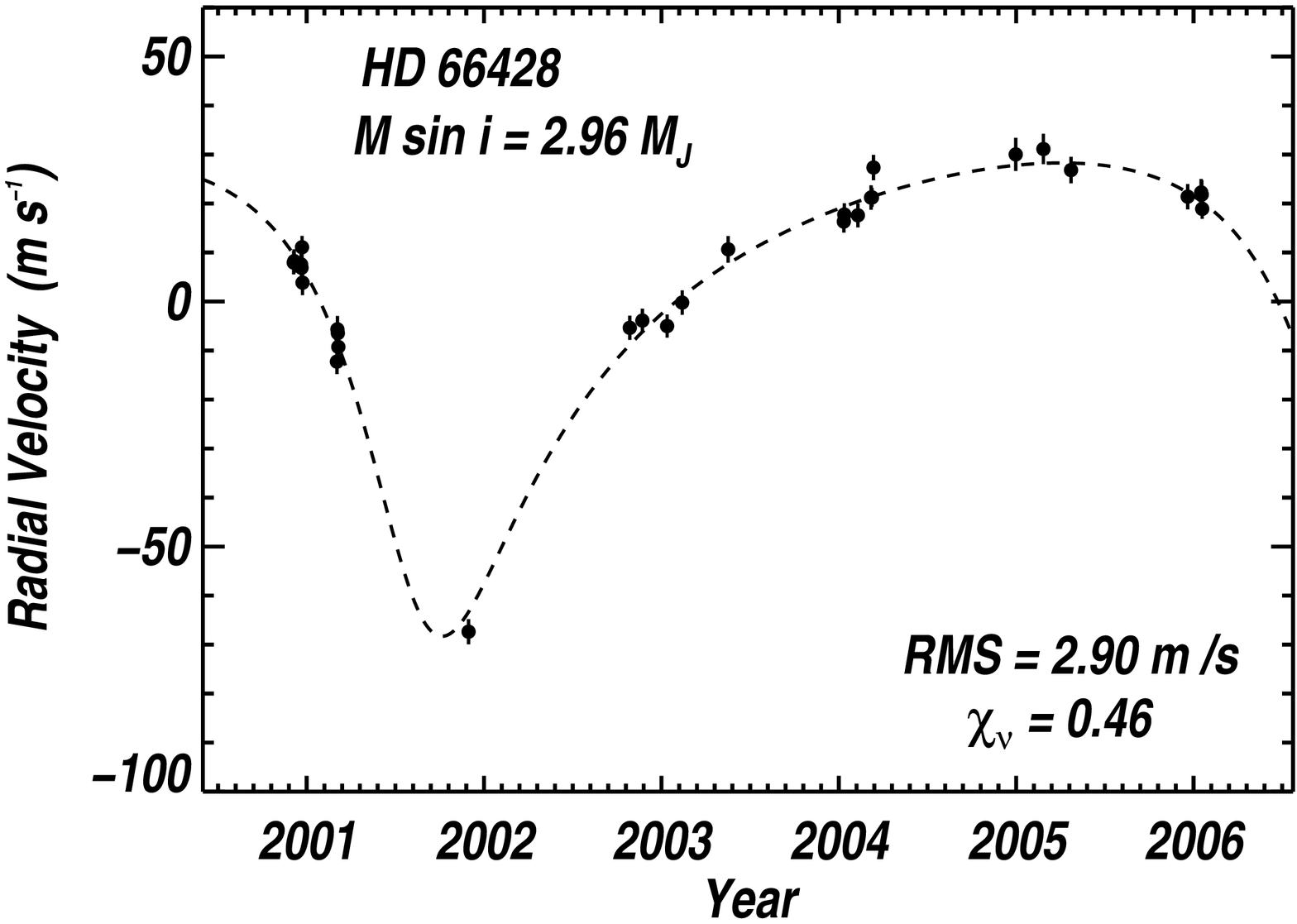}
\caption{\label{66428}Best-fit orbit to the radial velocities measured
  at Keck Observatory for HD 66428, with $P=5.4 yr$, $e=0.5$, and
  $\msini= 3 \mjup$.}
\end{figure}
\begin{figure}
\plotone{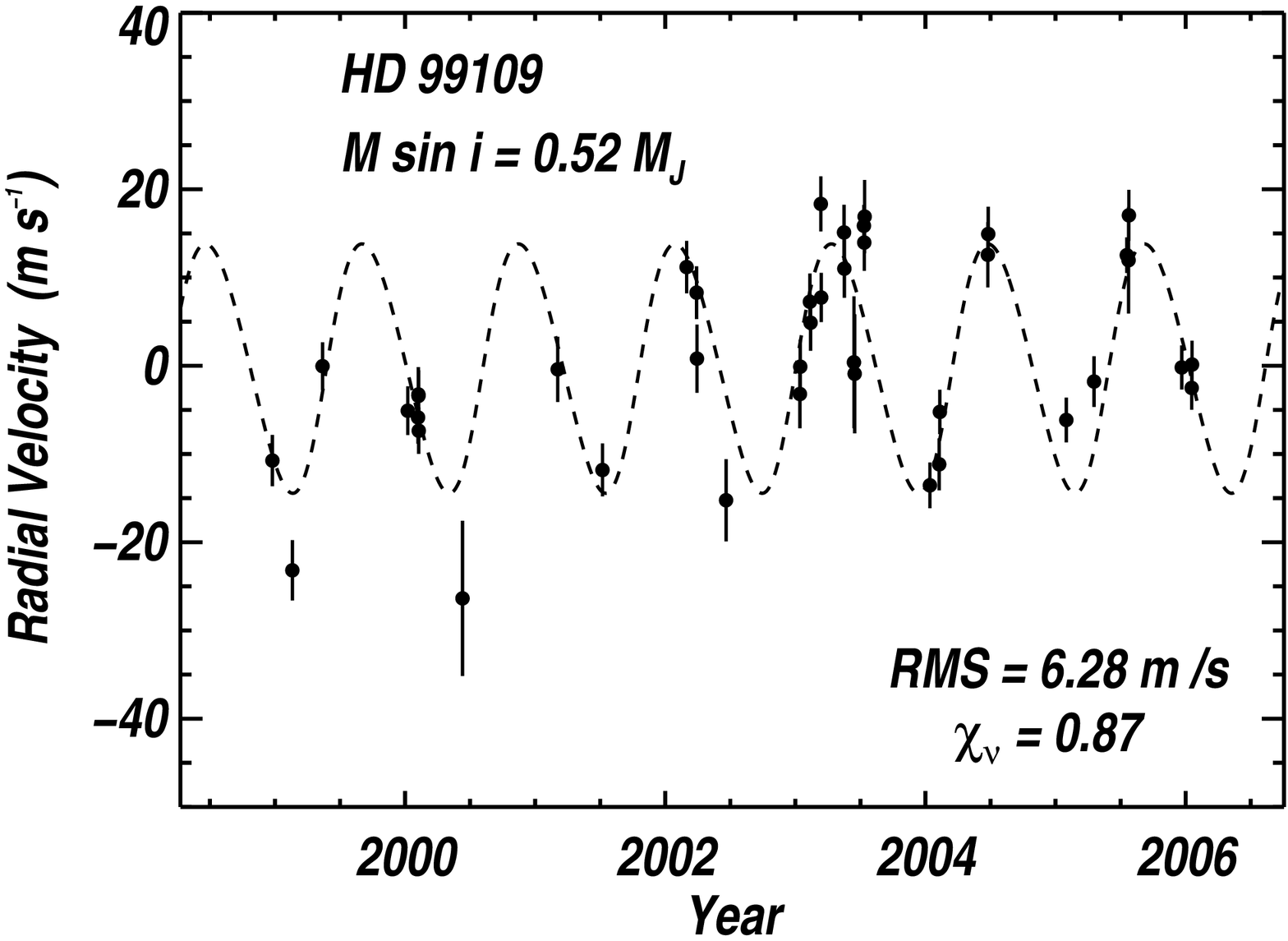}
\caption{\label{99109}Best-fit orbit to the radial velocities measured
  at Keck Observatory for HD 99109, with $P=1.2 yr$, $e \sim 0$, and
  $\msini= 0.5 \mjup$.}
\end{figure}
\begin{figure}
\plotone{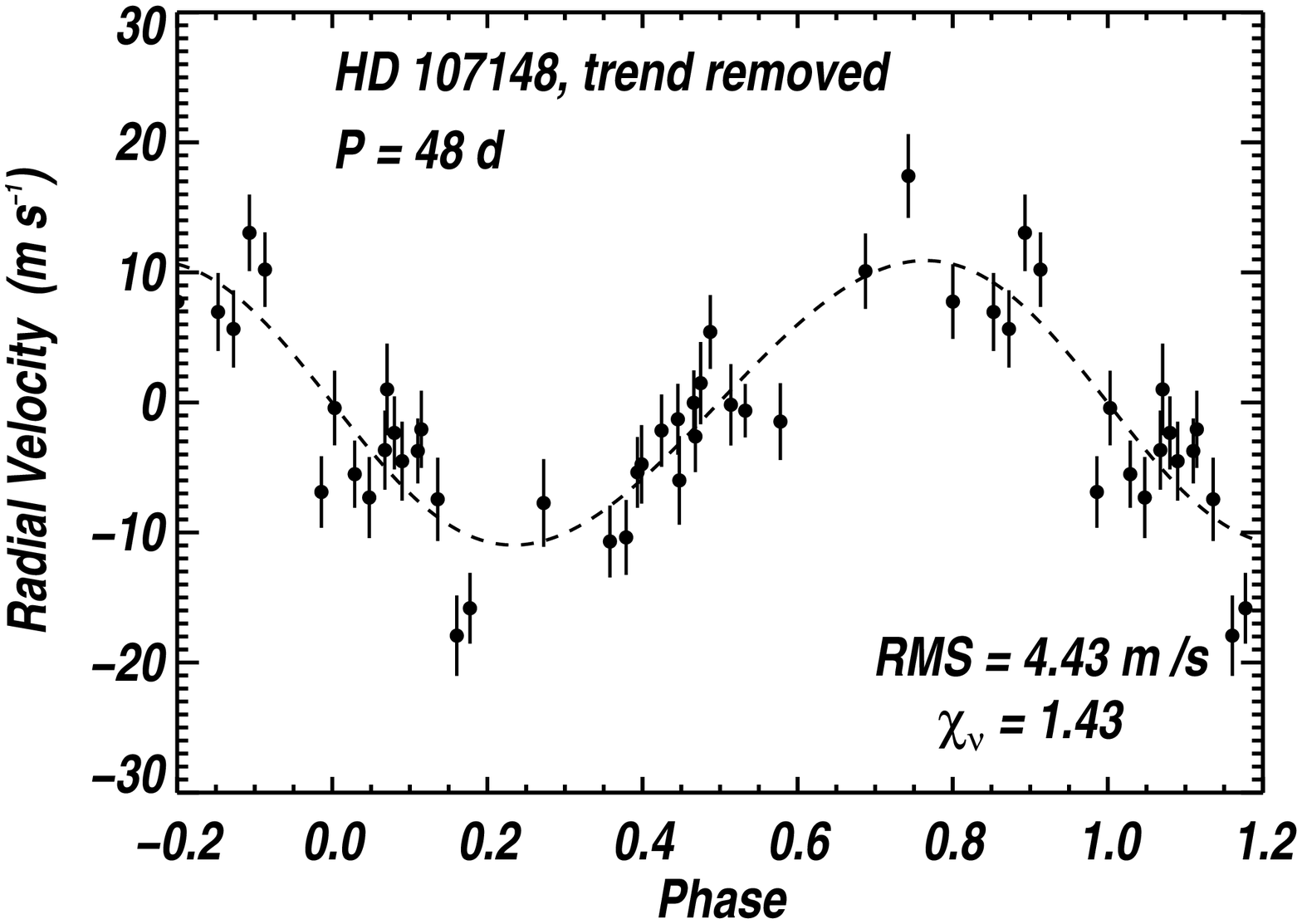}
\caption{\label{107148}Best-fit orbit to the radial velocities measured
  at Keck Observatory for HD 107148, with $P=48 d$, $e \sim 0$, and
  $\msini= 0.2 \mjup$.}
\end{figure}
\begin{figure}
\plotone{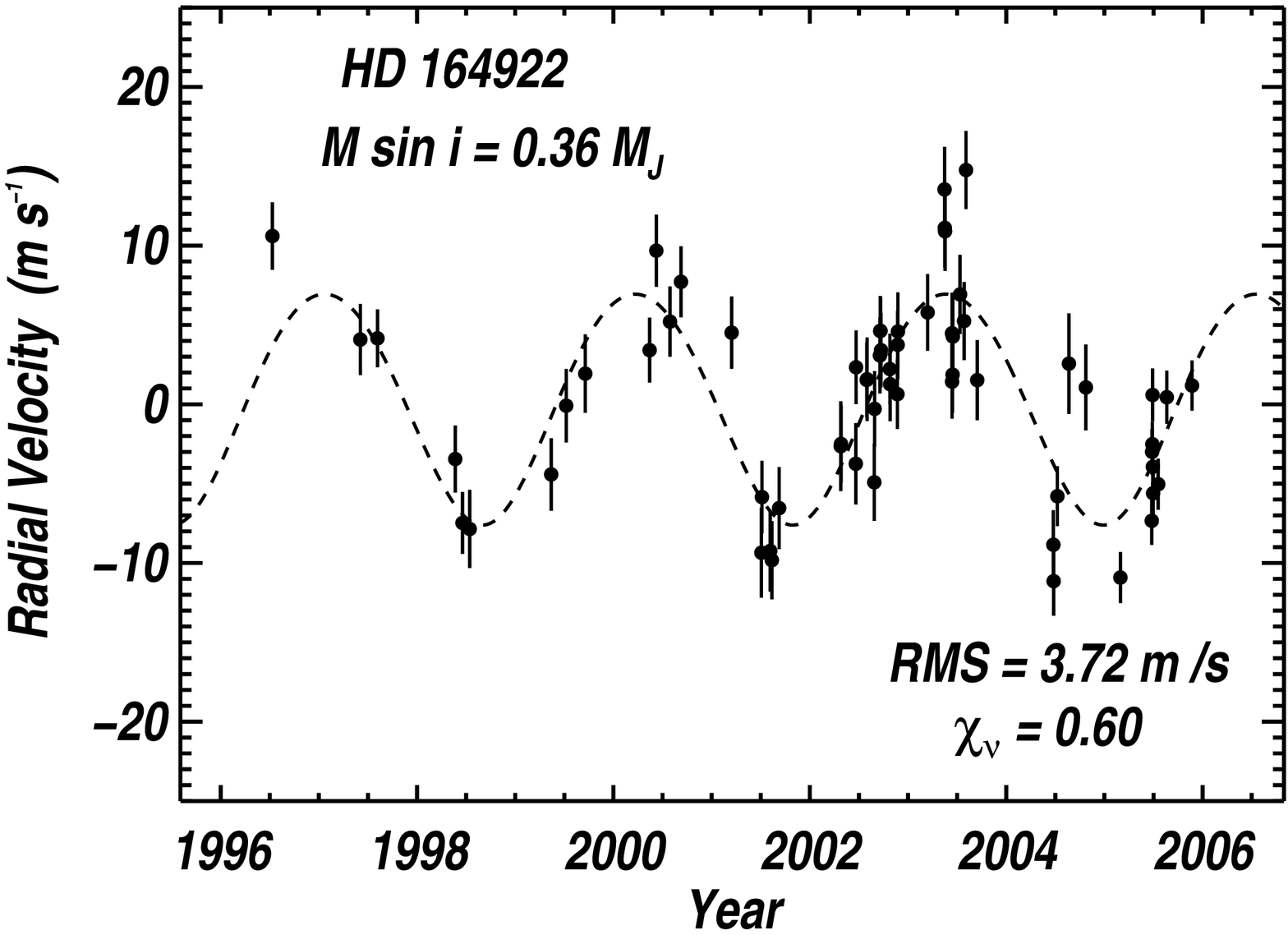}
\caption{\label{164922}Best-fit orbit to the radial velocities measured
  at Keck Observatory for HD 164922, with $P = 3.2 yr$, $e \sim 0$, and
  $\msini= 0.4 \mjup$.}
\end{figure}
\clearpage

\section{Discussion}
For many exoplanets we find an improved orbital solution when we
superimpose a linear trend and the velocity curve.  Such
systems likely 
contain additional companions of indeterminate mass and orbital
periods substantially longer than the span of the observations.  When
such systems are observed long enough that the radial velocity
signature of these more distant companions begin to deviate from a
linear trend, these fits naturally become poor, even though
double-Keplerian fits remain poorly constrained. 

An excellent example
is HD 13445 b, which shows a strong trend of $\sim -95$ m\persec,
consistent with the presence of a massive companion beyond 4 AU.  The
poor quality of the fit ($\sqrt{\chi^2_\nu} = 2.1$) may indicate
curvature in the signal of the massive companion --- indeed a
double-Keplerian fit with an outer planet with $P > 10$ yr produces a
fit with an r.m.s. of 4 m\persec.  This may be consistent with reports
of a massive companion at 20 AU \citep{Eggenberger03,Els01}.  A second
example is HD 68988, where the r.m.s. of the residuals of a
double-Keplerian are 3.3 m\persec, down from 6.4 m\persec\ for a
single Keplerian plus trend model.  In both of these cases the mass
and period of the more distant companion are under-constrained, so the
planetary nature of the companion is uncertain.  

A forthcoming work (Wright et al., 2006, in prep) will comb the
archive of velocities in Table~\ref{vels} for companions, such as HD
13345 c and HD 68988 c, of uncertain mass and orbital period.

{\it $\upsilon$ And} --- The precision of the Lick data prior to
1995 is not as high as today -- that pre-1995 data scatter about the fit
with an r. m. s. $\sim 100$ m \persec -- and data before 1992 are particularly
suspicious.  The orbital elements in the Table~\ref{orbittable}
represent a fit with data taken before 1992 excluded; Table~\ref{vels}
includes these pre-1992 data.

{\it HD 73526 b, c} --- These planets are in a 2:1 orbital resonance.
The dynamics of the system are discussed in \citet{Tinney06}.

{\it $\tau$ Boo b} --- The residuals to the fit of the 3.31 d planet
orbiting $\tau$ Boo show a trend of 15 m \persec y$^{-1}$ and may also
show some curvature.  The precision of the Lick data prior to 1995 is
not as high as it is today -- the fit for these times shows scatter of
$\sim 100$ m \persec -- and may not be reliable for constraining the
properties of the second companion.

{\it HD 149026 b} --- This planet transits its parent star.
\citet{Sato05} find $R=0.726\pm0.064 {\rm R}_{\rm Jup}$, and $i =
85.\!\!^{\arcdeg} 8^{+1.\!\!^{\arcdeg} 6}_{-1.\!\!^{\arcdeg} 3}$. The semi-amplitude, $K$, and
goodness-of-fit parameters in
Table~\ref{orbittable} represent the fit from data presented here, with
$P$ and $T_t$ held fixed at the values from \citet{Charbonneau06}.

{\it TrES-1} --- This planet transits its parent star, 2MASS
19040985+3637574 (GSC 02652-01324).
\citet{Alonso04} find $R = 1.08^{+0.18}_{-0.04} {\rm R}_{\rm Jup}$,
and $i = 88.\!\!^{\arcdeg} 5^{+1.\!\!^{\arcdeg} 5}_{-2.\!\!^{\arcdeg}
  2}$.  The semi-amplitude, $K$, and
goodness-of-fit parameters in
Table~\ref{orbittable} represent the fit from data presented here, with
$P$ and $T_t$ held fixed at the values from \citet{Alonso04}.

{\it HD 189733 b} --- This planet transits its parent star.
\citet{Bouchy05b} find $i = 85.\!\!^{\arcdeg} 3 \pm 0.1$ and $R = 1.26 \pm
0.03 {\rm R}_{\rm Jup}$.

{\it HD 209458 b} --- This planet transits its parent star.
\citet{Brown01} find $i = 86.\!\!^{\arcdeg} 1 \pm 0.\!\!^{\arcdeg} 1$
and $R = 1.347 \pm 0.06 {\rm R}_{\rm Jup}$ and \citet{Laughlin05b}
find an eccentricity consistent with 0.  The semi-amplitude, $K$, and
goodness-of-fit parameters in
Table~\ref{orbittable} represent the fit from data presented here, with
$P$ and $T_t$ held fixed at the values from \citet{Wittenmyer04}.

\section{Distribution of Exoplanets}

\begin{figure}
\caption{\label{mhist}Minimum mass distribution of the 167 known nearby
exoplanets with $\msini < 15$ AU.  The mass distribution shows a dramatic
decrease in the number of planets at high masses, a decrease that is
roughly characterized by a power law, $dN/dM \propto M^{-1.16}$.
Lower-mass planets have smaller Doppler amplitudes, so the relevent
selection effects enhance this effect.  This
distribution represents results from many surveys, and so is drawn
from an inhomogeneous sample.}
\plotone{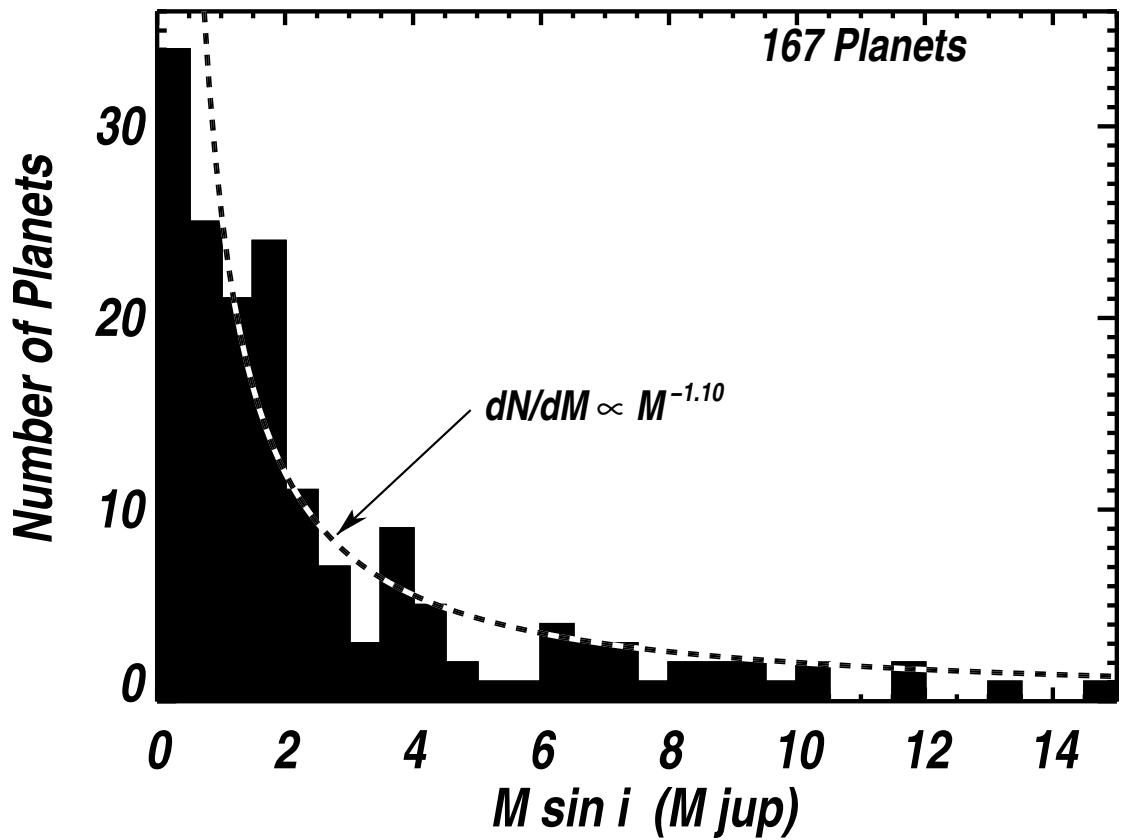}
\end{figure}

\begin{figure}
\caption{\label{ahist}Orbital distance distribution of the 167
known nearby exoplanets with $0.03 < a < 10$ in {\it logarithmic}
distance bins.  Planets with $a > 3 AU$ have periods
comparable to or longer than the length of most Doppler surveys, so
the distribution is incomplete beyond that distance.  This
distribution represents results from many surveys, and so is drawn
from an inhomogeneous sample.}
\plotone{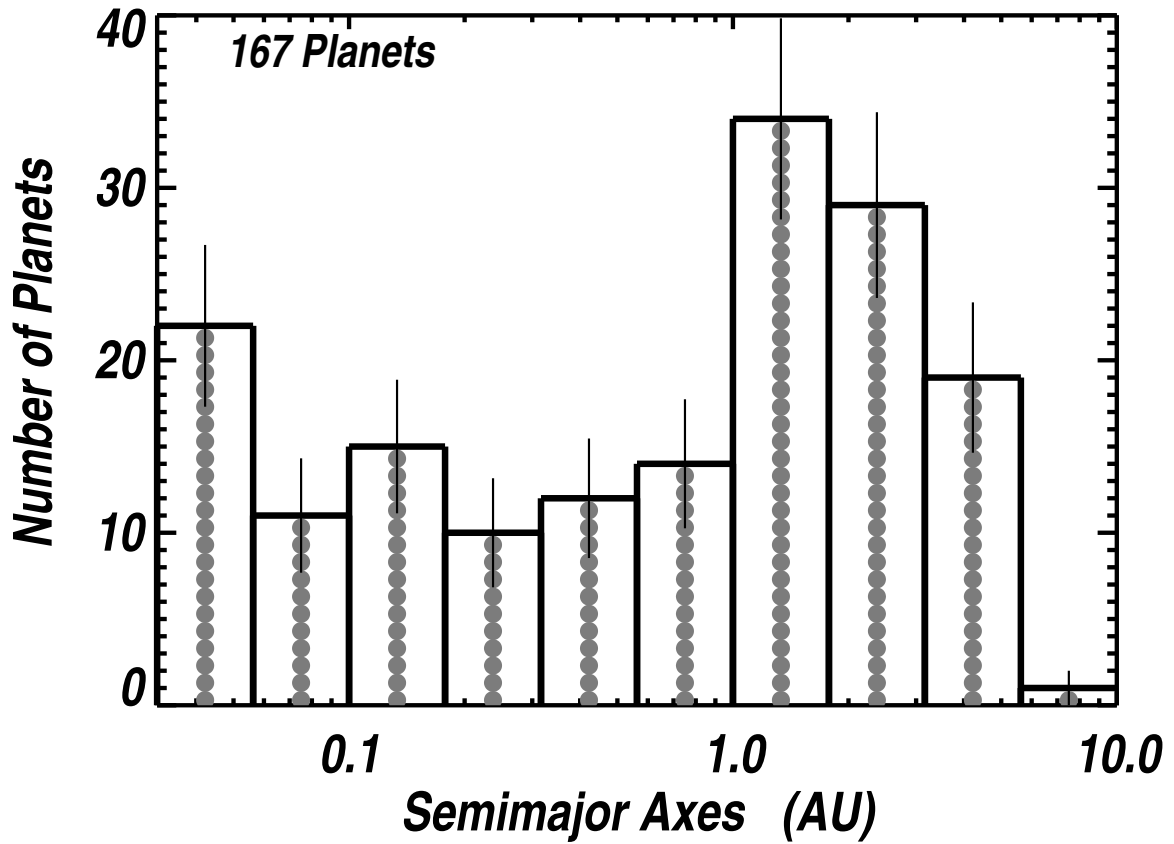}
\end{figure}

\begin{figure}
\caption{\label{phist}Distribution of periods among the known
nearby ``hot Jupiters''.  There is a clear ``pile-up'' of planets with
orbital periods near 3 days.  Doppler surveys generally have uniform
sensitivity to hot Jupiters, so for massive planets, there 
is no important selection effect contributing to the 3-day pile-up.  This
distribution represents results from many surveys, and so is drawn
from an inhomogeneous sample.}
\plotone{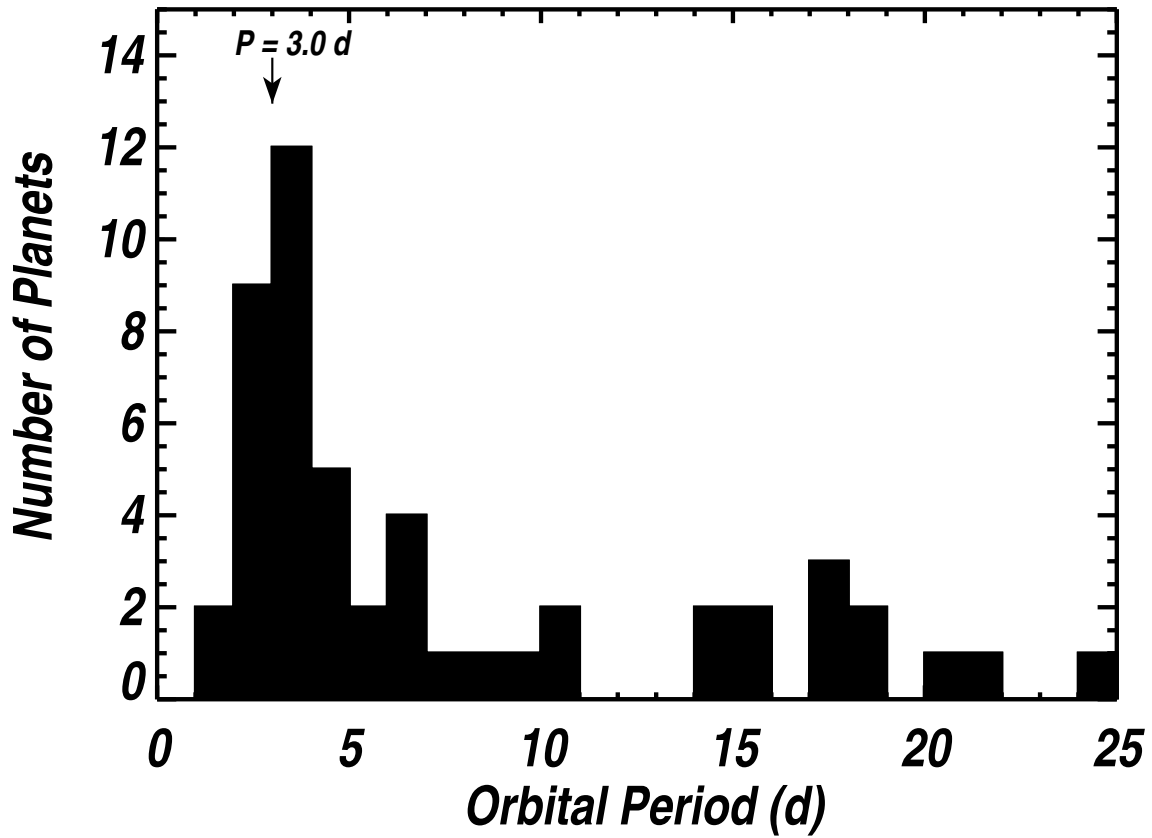}
\end{figure}

\begin{figure}
\caption{\label{mvsa}Minimum mass as a function
of semimajor axis for the 164 known nearby exoplanets with $0.03 < a <
6.5$ AU.  Doppler surveys are generally incomplete for exoplanets with
$a > 3$ AU, low-mass planets ($\msini < 1 \mjup$) beyond 1 AU, and
very low-mass planets ($\msini < 0.1 \mjup$) everywhere.  This
plot represents results from many surveys, and so is drawn
from an inhomogeneous sample.}
\plotone{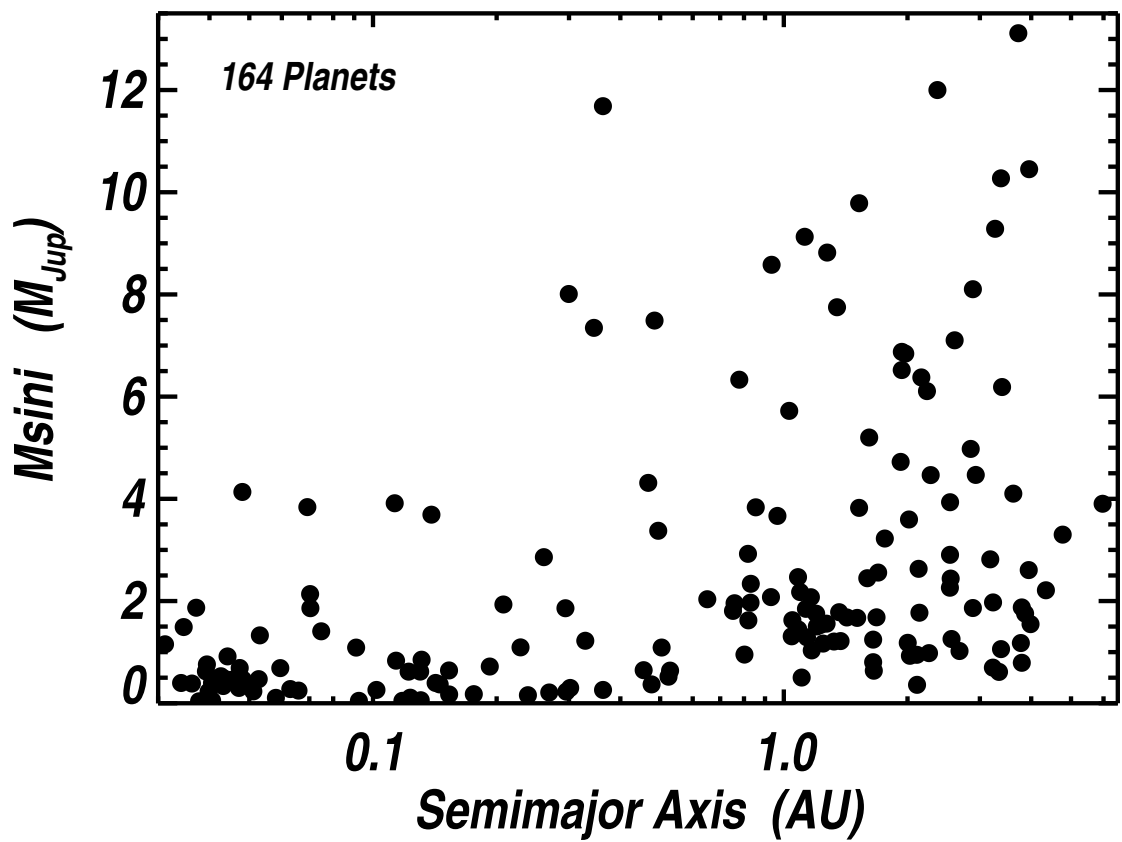}
\end{figure}

\begin{figure}
\caption{\label{evsa}Orbital eccentricity as a
function of semimajor axis for the 168 known nearby exoplanets.
Planets within 0.1 AU are presumably tidally circularized.  Beyond 0.1 AU, the
distribution of eccentricities appears essentially uniform between 0
and 0.8.  For most Doppler surveys, sensitivity is not a strong
function of eccentricity for $0 < e < 0.8$ and $a < 3$ AU.  This
plot represents results from many surveys, and so is drawn
from an inhomogeneous sample.}
\plotone{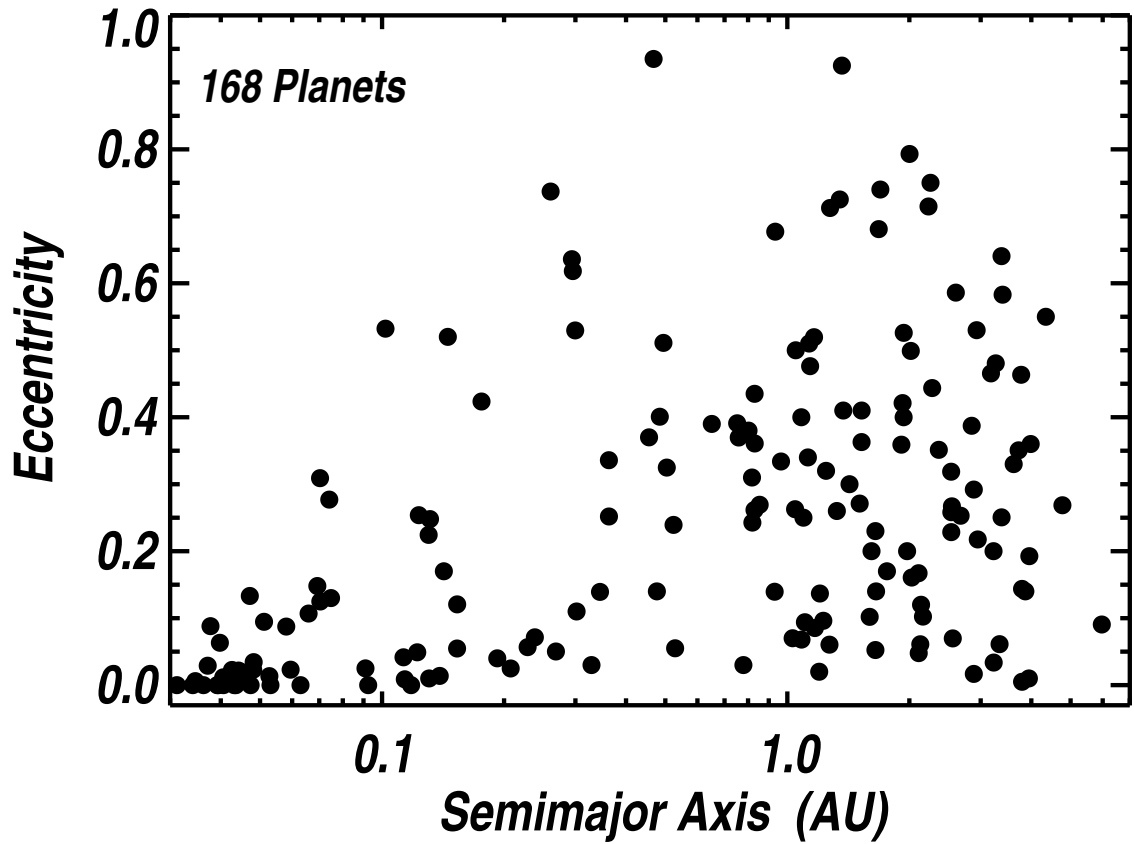}
\end{figure}

\begin{figure}
\caption{\label{evsm}Distribution of orbital eccentricities as a
function of minimum mass for the 130 known nearby exoplanets with $\msini
< 13 \mjup$, excluding those for which $a < 0.1$ AU, i.e., those
planets which may have been tidally circularized.  High-mass
exoplanets ($\msini > 5 \mjup$) have a slightly higher median
eccentricity than lower-mass exoplanets. The
completeness of Doppler surveys increases with \msini\ and is
generally insensitive to eccentricity.  This
distribution represents results from many surveys, and so is drawn
from an inhomogeneous sample.}
\plotone{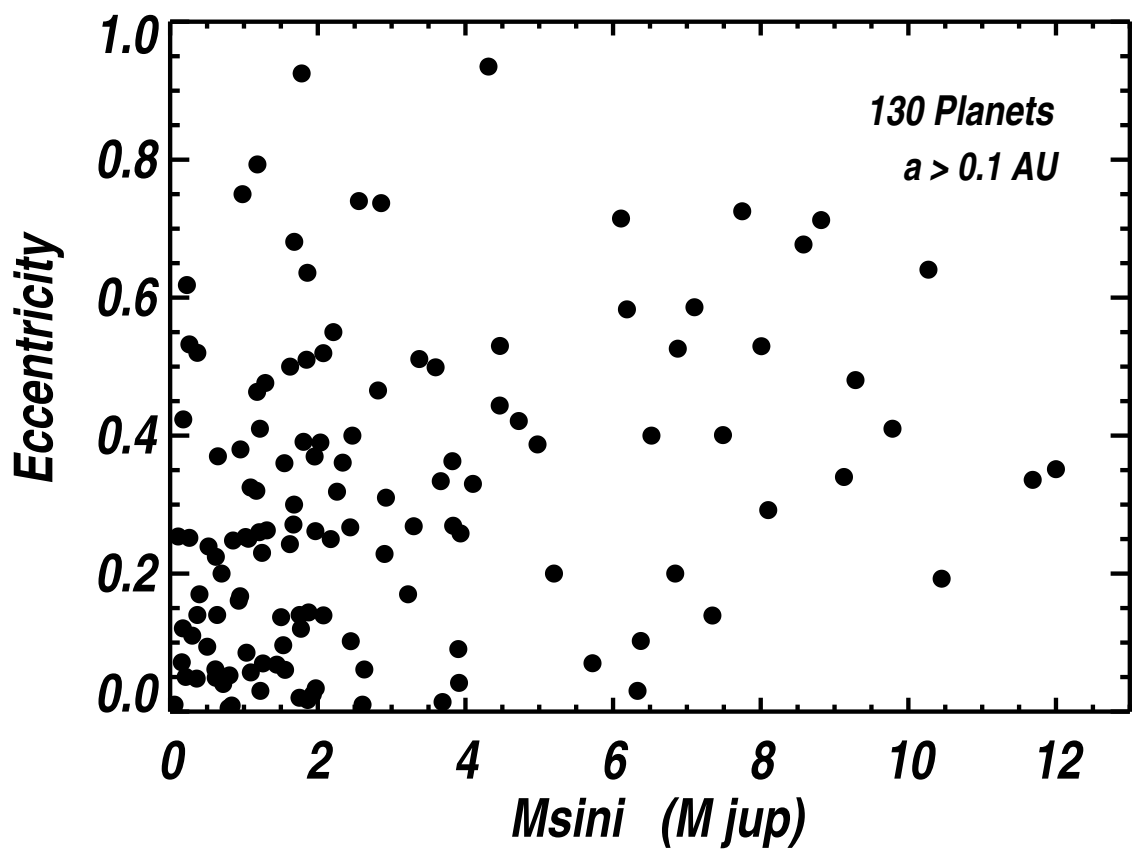}
\end{figure}

Fig.~\ref{mhist}-\ref{evsm}
show the distribution of the exoplanets in this catalog.  One must
take care when interpreting these figures for at least two reasons:
firstly, selection effects make some aspects of these distributions
inconsistent with the parent population of exoplanets, and 
secondly, the selection effects of the various planet search programs
are different. \citet{Butler05} and \citet{Marcy_Japan_05} analyze the properties and distribution of
planets detected around 1330 FGKM dwarfs monitored at Lick, Keck, and
the AAT, and discuss the biases in and uniformity of that sample.  The
figures presented here are best interpreted as describing the
distribution of properties of the known exoplanets as drawn from
multiple, nonuniform samples, as opposed to that of the parent
population of exoplanets. 

The target list for the California, Carnegie, and Anglo-Australian
Planet Searches has been  
published in \citet{Wright04}, \citet{Nidever02}, and \citet{Jones02}.
A complete target list for the Geneva group is not public and not
recoverable, though a list of HARPS target stars is presently
available on the ESO
website\footnote{http://www.eso.org/observing/proposals/gto/harps/}.   
Both searches may be considered roughly magnitude
limited within a set of \bv\ bins excluding giant stars, but both
groups have also added additional stars using other criteria (such as
metallicity).

Fig.~\ref{mhist} shows the minimum mass distribution of the 167 known nearby
exoplanets with $\msini < 15$ AU.  The mass distribution shows a dramatic
decrease in the number of planets at high masses, a decrease that is
roughly characterized by a power law, $dN/dM \propto M^{-1.1}$, affected
very little by the unknown $\sin{i}$ \citep{Jorissen01}.  We have
calculated the exponent in this power law with a linear least-squares
fit to the logarithm of the mass distribution assuming Poisson
errors.  We neglected uncertainties in the masses of the planets due
to uncertainties in stellar masses and the $\sin{i}$ ambiguity.  For
this reason, and because the surveys that detected these planets have
heterogeneous selection effects, we regard this power law simply as a rough
description of the distribution of known planets.  \citet{Cumming06}
finds, for the more uniform sample of the California and Carnegie
Planet Search, that the distribution of planets with $P > 100$ days is
well-fitted with a broken power law:

\begin{equation}
dN/dM \propto \left\{ \begin{array}{ll}
               M^{-1.2} & M < 0.6 \mjup \\
               M^{-1.9} & M > 0.6 \mjup \\
                     \end{array} \right.
\end{equation}

The low end of this distribution suffers from a selection effect common to all
Doppler surveys: low-mass planets induce small velocity variations, so
are difficult to detect and under-represented in Fig.~\ref{mhist}.
Massive planets are easier to detect, making the apparent paucity of planets
with $M >3 \mjup$, and that of objects with $M >12 \mjup$ (the
``brown dwarf desert'') real.

Fig.~\ref{ahist} shows the orbital distance distribution of the 167
known nearby exoplanets with $0.03 < a < 10$.  Since orbital distance
is a function of orbital period, the existing Doppler surveys are
increasingly incomplete for $a \gtrsim 3$ AU, corresponding to $P
\gtrsim 5$ years.  Note that the abscissa is logarithmic. Among the
1330 FGKM dwarfs studied by \citet{Marcy_Japan_05}, the 
occurrence rate of planets within 0.1 AU is 1.2\%.  A modest (flat)
extrapolation beyond 3 AU (in logarithmic bins) suggests that there
exist roughly as many planets at distances between between 3-30 AU as
below 3 AU, making the occurrence of giant planets roughly 12\% within
30 AU.  The rapid rise of planet frequency with semi-major axis beyond
0.5 AU portends a large population of Jupiter-like planets beyond 3 AU.

Fig.~\ref{phist} shows the distribution of periods among the known
nearby ``hot Jupiters''.  There is a clear ``pile-up'' of planets with
orbital periods near 3 days, suggesting that whatever orbital
migration mechanism brings these giant planets close to their parent
stars ceases when they reach this period.  Alternatively, some
breaking meachanism stops them there, or weakens inward of the
distance, sending the planets into the star.  Note that the Doppler
surveys generally have uniform sensitivity to hot Jupiters at all of
the orbital periods in Fig.~\ref{phist}, so for massive planets there
is no important selection effect contributing to the 3-day pile-up.

Fig.~\ref{mvsa} shows minimum mass as a function
of semimajor axis for the 164 known nearby exoplanets with $0.03 < a <
6.5$ AU.  There is a dearth of close-in exoplanets with high mass which
cannot be due to a selection effect since high-mass planets have large
Doppler signatures -- indeed Doppler surveys are generally complete
with respect to high-mass, close-in exoplanets.  Selection effects 
make detection of low-mass planets beyond 1 AU difficult, however, so
it is not clear that the mass distribution for planets beyond 1 AU is
different from that of hot Jupiters.

Fig.~\ref{evsa} shows orbital eccentricity as a
function of semimajor axis for 168 known nearby exoplanets.
Planets within 0.1 AU are nearly always on circular or nearly circular orbits,
presumably due to tidal circularization.  Beyond 0.3 AU, the
distribution of eccentricities appears essentially uniform between 0
and 0.8.  For most Doppler surveys, sensitivity is not a strong
function of eccentricity for $0 < e < 0.7$ and $a < 3$ AU.

Fig.~\ref{evsm} shows orbital eccentricity as a
function of minimum mass for nearby exoplanets with $\msini
< 13 \mjup$.  We have excluded those planets which may have been
tidally circularized, i.e. those for which $a < 0.1$ AU.  This
figure shows no strong correlation between eccentricity and mass,
but close inspection shows that high-mass exoplanets ($\msini > 5 \mjup$)
have a higher median eccentricity than lower-mass exoplanets.  The
completeness of Doppler surveys increases with \msini\ and is
generally insensitive to eccentricity for $e < 0.7$.

\section{Conclusions}

We have remeasured precise orbital elements for planets orbiting stars for which we have precision radial velocity data
from Keck, Lick, and AAO using the latest data and
improved data reduction techniques.  In
addition, we have compiled the published orbital 
parameters of all other exoplanets within 200 pc, as well as
spectroscopically-derived stellar parameters of their host stars.
Finally, we present four new extrasolar planets, bringing to 172 the total of
known exoplanets in this catalog with a minimum mass $\msini < 24
\mjup$.

The 172 known exoplanets span a range of eccentricities, which weakly
correlate with minimum planetary mass.  Planets within 0.1 AU are nearly
always in circular orbits, presumably due to tidal circularization. 
The 3-day ``pile-up'' and the ``brown dwarf desert'' are both strongly
apparent and unaffected by the important observational biases.
Finally, the mass distribution increases sharply toward lower
masses (roughly as the inverse of the minimum planetary mass) and
toward higher orbital distance.  Since these regions are where current
surveys are most incomplete, this implies that many more low-mass
planets and long-period await discovery as Doppler surveys cover a
longer time baseline and become more precise.  A forthcoming work will
discuss some more speculative exoplanet candidates of this nature just
emereging in from our planet searches.

\clearpage

\clearpage
\pagestyle{empty}
\begin{deluxetable}{lllrrcccllrclllrc}
\rotate
\tabletypesize{\tiny}
\tablecolumns{17}
\tablewidth{0pc} 
\tablecaption{Properties of Exoplanet Host Stars\label{proptable}}
\notetoeditor{I feel that for the first entry on each page of this
  table, RA and Dec fields should be annotated with
 h, m, \fs, \arcdeg, \arcmin, and \farcs.  I'm not sure how to
 implement this without knowing in advance how many lines will appear on each
 page in the final version}
\tablehead{
\colhead{HD}&\colhead{Hip
  \#}&\colhead{Alt. name}&\colhead{RA}&\colhead{Dec.}&\colhead{\bv}&\colhead{$V$}&\colhead{Distance}&\colhead{\teff}&\colhead{$\log{g}$}&\colhead{[Fe/H]}&\colhead{$v\sin{i}$}&\colhead{Mass}&\colhead{ref.\tablenotemark{a}}&\colhead{S}&\colhead{\dmv}&\colhead{jitter}\\
            & 
&     &  \colhead{(J2000)}     &  \colhead{(J2000)}             &             && \colhead{(pc)}
  &\colhead{(K)}       &    \colhead{(cm $\mbox{s}^{-2}$)}   &
& \colhead{$km \persec$}
&  \colhead{(\Msol)} & & & &\colhead{(m \persec)}}
\startdata
142 & 522 &  & 00 06 19.176   & -49 04 30.69 & 0.52 & 5.70 & 25.64(42) & 6249 & 4.185 & 0.100 & 10.35 & 1.24 & VF5 & 0.16 & 0.28 & 4.3\\
1237 & 1292 & GJ 3021 & 00 16 12.677   & -79 51 04.25 & 0.75 & 6.59 & 17.62(16) & 5536 & 4.56 & 0.120 & \nodata & 0.90 & Sn4 & \nodata & 0.03 & \nodata\\
2039 & 1931 &  & 00 24 20.278   & -56 39 00.17 & 0.66 & 9.00 & 89.8(9.1) & 5941 & 4.384 & 0.315 & 3.25 & 1.17 & VF5 & 0.18 & 0.60 & 3.7\\
2638 & 2350 &  & 00 29 59.872   & -05 45 50.41 & 0.89 & 9.44 & 53.7(3.9) & 5192 & 4.29 & 0.160 & \nodata & 0.93 & Sn5 & \nodata & 0.31 & \nodata\\
3651 & 3093 & 54 Psc & 00 39 21.806   & +21 15 01.70 & 0.85 & 5.88 & 11.107(89) & 5221 & 4.453 & 0.164 & 1.15 & 0.89 & VF5 & 0.17 & 0.28 & 3.5\\[5pt]
4208 & 3479 &  & 00 44 26.650   & -26 30 56.45 & 0.66 & 7.78 & 32.7(1.2) & 5600 & 4.517 & -0.284 & \nodata & 0.87 & VF5 & 0.17 & -0.33 & 3.7\\
4308 & 3497 &  & 00 44 39.268   & -65 38 58.28 & 0.65 & 6.55 & 21.85(27) & 5695 & 4.580 & -0.310 & \nodata & 0.90 & VF5 & \nodata & -0.03 & \nodata\\
4203 & 3502 &  & 00 44 41.202   & +20 26 56.14 & 0.77 & 8.70 & 77.8(7.7) & 5702 & 4.361 & 0.453 & 1.23 & 1.13 & VF5 & 0.14 & 1.27 & 4.0\\
\enddata
\tablenotetext{a}{References are encoded as follows:
Ad9: \citet{Apl99}; Bc5:
  \citet{Bouchy05b}; Bu4: \citet{Butler04}; Da6: \citet{DaSilva06};  Fi5: \citet{Fischer05};
  Fi6: \citet{Fischer06}; Wr6: J. T. Wright et al. (2006) in prep.;
Fu4: \citet{Fuhrmann04}; Ge5: \citet{Ge05}, http://vo.obspm.fr/exoplanets/encyclo/planet.php?p1=HD+102195\&p2=b; Jh6: J. A. Johnson et al. 2006; Mc8:
\citet{Marcy98}; Nd4: \citet{Nordstrom04}; 
Sn4: \citet{Santos04a};  Sn5: \citet{Santos05}; 
St3: \citet{Sato03};  St5: \citet{Sato05}; Sw3: \citet{Setiawan03}; 
Sz4: \citet{Sozzetti04}; Sz6: \citet{Sozzetti06}; VF5: \citet{Valenti05}; Z3: \citet{Zucker03}}

\tablenotetext{b}{This giant star has a poorly determined mass;  both
  estimates are plausible.}
\tablenotetext{c}{No effective temperature available: \teff\ is
  estimated from \bv\ using \citet{Flower96}.}
\tablenotetext{d}{No $S$-value available: star is assumed to be very
  inactive.}
\tablenotetext{e}{BD -10\arcdeg3166 appears to be a metal-rich K0 dwarf.
  Based upon its spectroscopic similarity to 55 Cnc \citep{Gonzalez99}, a star 4.1 mag
  brighter in $V$ band, we can crudely infer a distance $\sim$ 7 times greater than that
  star, or $\sim$ 80 pc.}
\tablenotetext{f}{No parallax available: star is assumed to be on
  the main sequence.}
\end{deluxetable}
\clearpage
\pagestyle{plaintop}
\begin{deluxetable}{rl@{ }rcccccccccccrl}
\notetoeditor{The spacings in this table are problematic.  My
  intention is for planets within a system to have entries more
  closely spaced (to emphasize their related nature), and not to wrap
  between pages (since the star name and fit quality are not repeated
  for each component).  In addition, I would like to add wider
  spacings between every fifth entry per page to aid the eye.  Ideally,
  there would be 40 entries per page, but the duplicate entry at HD
  37124 throws the system off.  I think things would work if I could
  somehow squeeze entry 40 onto the first page to keep the HD 38529 planetary
  system together and thus keep 40 per page for the rest of the table.
  Another issue is that superscripts in the table tend to play with
  the spacings between the rows, making these distinctions harder to
  make out.  I would prefer that the spacings not be affected by the
  presence of superscripts (most visible, perhaps, between entries 49
  and 50 -- the 'b' note for eccentricity throws off the spacing.)} 
\rotate
\tabletypesize{\tiny}
\tablecolumns{16}
\tablewidth{0pc} 
\tablecaption{Catalog of Nearby Extrasolar Planets\label{orbittable}}
\tablehead{
&\multicolumn{2}{c}{Planet}  &\colhead{Per}              &\colhead{K}&\colhead{e}
&\colhead{$\omega$}          &\colhead{$T_{\mbox{p}}$}   &\colhead{$T_{\mbox{t}}$} &\colhead{trend}        
&\colhead{\msini}            &\colhead{a}                &\colhead{r.m.s.}
&\colhead{$\sqrt{\chi^2_\nu}$}&\colhead{N$_{\mbox{obs}}$}&\colhead{ref.\tablenotemark{a}(alt.)}\\ 
&                            &       &\colhead{(d)}&\colhead{(\mps)}&
& \colhead{($\deg$)}&\colhead{(JD-2440000)}&\colhead{(JD-2440000)}&\colhead{(\mps yr$^{-1}$)}
&\colhead{($\Mjupsmall$)} &\colhead{(AU)}&\colhead{(\mps)}&
                      &                          &
}
\startdata
       1 & HD 142 & b & 350.3(3.6) & 33.9(4.7) & 0.26(18)\tablenotemark{b} & 303\tablenotemark{b} & 11963(43) & 11737(25) & -10.4(1.1) & 1.31(18) & 1.045(61) & 12 & 1.5 & 53 & Bu6\\[1pt]
       2 & HD 1237 & b & 133.71(20) & 167.0(4.0) & 0.511(17) & 290.7(3.0) & 11545.86(64) & \nodata &  & 3.37(49) & 0.495(29) & 19 & 1.8 & 61 & Nf1b\\[1pt]
       3 & HD 2039 & b & 1120(23) & 153(22) & 0.715(46) & 344.1(3.6) & 12041(13) & 10992(26) & 3.5(1.5) & 6.11(82) & 2.23(13) & 11 & 0.84 & 41 & Bu6\\[1pt]
       4 & HD 2638 & b & 3.44420(20) & 67.40(40) & 0\tablenotemark{c} & 0\tablenotemark{c} & 13323.2060(20) & \nodata &  & 0.477(68) & 0.0436(25) & 3.3 & \nodata & 28 & Mo5\\[1pt]
       5 & 54 Psc & b & 62.206(21) & 16.0(1.2) & 0.618(51) & 233.3(7.4) & 12189.83(68) & 12176.3(1.9) &  & 0.227(23) & 0.296(17) & 6.6 & 1.3 & 163 & Bu6\\[8pt]
       6 & HD 4208 & b & 828.0(8.1) & 19.06(73) & 0.052(40)\tablenotemark{b} & 345\tablenotemark{b} & 11040(120) & 10440(16) &  & 0.804(73) & 1.650(96) & 3.4 & 0.72 & 41 & Bu6\\[1pt]
       7 & HD 4308 & b & 15.560(20) & 4.07(20) & 0.000(10) & 359(47) & 13311.7(2.0) & \nodata &  & 0.0467(70) & 0.1179(68) & 1.3 & 1.4 & 41 & Udry 2005\\[1pt]
       8 & HD 4203 & b & 431.88(85) & 60.3(2.2) & 0.519(27) & 329.1(3.1) & 11918.9(2.7) & 11558.7(7.2) & -4.38(71) & 2.07(18) & 1.164(67) & 4.1 & 0.80 & 23 & Bu6\\
\enddata
\end{deluxetable}

\begin{deluxetable}{rl@{ }rcccccccccrl}
  \rotate
  \tabletypesize{\tiny}
  \tablecolumns{14}
\tablewidth{0pc} 
\tablecaption{Independent Orbital Solutions\label{alttable}}
\tablehead{
\multicolumn{2}{c}{Planet}  &\colhead{Per}              &\colhead{K}&\colhead{e}
&\colhead{$\omega$}          &\colhead{$T_{\mbox{p}}$}   &\colhead{trend}        
&\colhead{\msini}            &\colhead{a}                &\colhead{r.m.s.}
&\colhead{$\sqrt{\chi^2_\nu}$}&\colhead{N$_{\mbox{obs}}$}&\colhead{ref.\tablenotemark{a}}\\ 
                             &       &\colhead{(d)}&\colhead{(\mps)}&
& \colhead{($\deg$)}&\colhead{(JD-2440000)}&\colhead{(\mps y$^{-1}$)}
&\colhead{($\Mjupsmall$)} &\colhead{(AU)}&\colhead{(\mps)}&
                      &                          &
}
\startdata
HD 8574 & b & 277.55(77) & 66(5) & 0.288(53) & 3.6(10.9) &
111467.5(6.6) & & 2.11 & 0.77 & 13 & 1.4 &
41 & Pr3 \\

$\upsilon$ And & b & 4.61712(9) & 77.2(1.3) & 0.02(23) & 242(37) &
10004.28(48) & & 0.75(1) & 0.059 & 15 & 2.1 &
71 & Nf4 \\
             & c & 238.10(46) & 63.0(1.7) & 0.185(28) & 214(11) &
10159.4(8.0) &         & 2.25(6) & 0.821 &     &    
   &  & \\
             & d & 1319(18)   & 63.8(2.3) & 0.269(36) & 248(11) &
 9963(53)   &         & 3.95(13) & 2.57 &     &    
   & & \\

HD 10647     & b & 1040(37)   & 18(1)     & 0.18(8)   & 68(17) &
12261(47)    & & 0.91     & 2.1      & 8.2 &     &
72 & My3\\

HD 13445     & b & 15.78(4)   & 380(1)    & 0.046(4)  & 270(4) & 
11146.7(2)   & -131    &    4 & 0.11     & 7   &     & 
61 & Q0\\

$\iota$ Hor   & b & 311.3(1.3) & 68(4)     & 0.22(6)   & 79(13) & 
11309(10)& & 2.24(13) & 0.91     & 23.2 & 1.5 &
88 & Nf1b\\

$\epsilon$ Eri &b & 2502(20) & 19.0(1.7)   & 0.608(41) & 48.9(4.1)&
9195(14)     & 0.42(20) & 0.86    &          & 14   &     &
225 & H0\\
\enddata
\tablenotetext{a}{References are encoded in Table~\ref{orbittable}.}
\tablenotetext{b}{Parameter held fixed in fit.}
\end{deluxetable}
\clearpage
\acknowledgments
The authors wish to thank the many observers over many years who
helped gather the data herein at telescopes around the world, and the
many collaborators who helped reduce, analyze and interpret this
inestimable data set, including Jeff Valenti, Bernie Walp, Andrew Cumming, Kevin
Apps, Eugenio Rivera, Greg Laughlin, Sabine Frink, Tony Misch, Grant
Hill, David Nidever, Eric Nielsen, Amy Reines, Joe Barranco, Bob
Noyes, Eric Williams, Preet Dosanjh, Mike Eiklenborg, Mario Savio,
Heather Hauser, and Barbara Schaefer.

The authors wish to recognize and acknowledge the
very significant cultural role and reverence that the summit of Mauna
Kea has always had within the indigenous Hawaiian community.  We are
most fortunate to have the opportunity to conduct observations from
this mountain.

This research has made use of the SIMBAD database, operated at CDS,
Strasbourg, France, and of NASA's Astrophysics Data System
Bibliographic Services, and is made possible by the generous support
of Sun Microsystems, NASA, and the NSF.

\end{document}